\documentclass[superscriptaddress,aps,prl,amsfonts,twocolumn]{revtex4-1}
\usepackage{epsfig}
\usepackage{textcomp}
\begin{document}
\title{The entrire history of a photon}
\author{Marcin Wie\'sniak}
\affiliation{Institute of Informatics, Faculty of Mathematics, Physics, and Informatics, University of
Gda\'{n}sk, PL-80-308 Gda\'{n}sk, Poland}
\begin{abstract}
Using the most basic mathematical tools, I present the full analysis of the experiment described in [A. Danan, D. Farfurnik, S. Bar-Ad, and L. Vaidman, {\em Phys. Rev. Lett.} {\bf 111}, 240402 (2013)]. First, I confirm that the data presented therein are in full agreement with the standard quantum mechanics. I then show other symptoms of presence of photons at all mirrors in the setup. I then analytically explain both the absence of peaks a Reader of [A. Danan, D. Farfurnik, S. Bar-Ad, and L. Vaidman, {\em Phys. Rev. Lett.} {\bf 111}, 240402 (2013)] is made to expect, and presence of those not discussed in the Reference.
\end{abstract}
\maketitle
Understanding the nature of light, being the basic source of our knowledge about the World and the entire Universe, has always been one of the most appealing philosophical and physical problems. Significant contributions over time were made by, among others, Empodocles, Leucippus, Aristotle, Newton, Huygens, Fresnel, and Einstein. The final conclusion is that it propagates as a wave, but causes effects (i.e. is being detected) like a particle. This duality was the key experimental evidence for quantum mechanics (also called wave mechanics), which combines smooth wave-like propagation of probability amplitudes with sharp projections of a state due to measurements.

An intriguing feature of light was predicted already in 1801 by Young and demonstrated two years later in the double-slit experiment \cite{YOUNG}, to be later simplified to the use of a Mach-Zehnder interferometer (MZI) \cite{MZI}. The idea of the experiment is to give light two separate paths of distribution, which are later recombined. The emerging pattern is not a sum of images generated by light passing through individual slits or paths. Instead, we observe fringes dependent on wavelengths of the propagating wave, which are precisely explained by quantum mechanics. Wheeler's proposal of the delayed choice experiment \cite{WHEELER}, put to test in 2007 \cite{ASPECT}, shows that an observer can freely choose between wave-like and particle-like propreties of photons, even inside an interferometer.

More recently, a nested three-path Mach-Zehnder interferometer was proposed \cite{VAID1} and realized by Danan, Farfunik, Bar-Ad, and Vaidman \cite{VAIDMAN}. In one (upper) path of a MZI there are two mirrors bending the beam and another (small) MZI, one input and one output of which are combined with the upper path. A mirror in the lower left corner of the large interferometer is labeled $C$, the corners of the small interferometer are denoted by $A$ and $B$, while the mirrors in front of and behind the small loops are $E$ and $F$. In the experiment, each of these mirrors vibrated with each own frequency $f_A, ..., f_F$ and interferometer was tuned for destructive interference (See Fig. 2 in Ref \cite{VAIDMAN}). The experiment's outcome is the power spectrum of the light reaching the detector. Quite surprisingly, the spectrum is peaked at $f_A, f_B,$ and $f_C$, but not at $f_E$ or $f_F$. The bold interpretation of this fact is expressed already in the title of Ref. \cite{VAIDMAN}: ``Asking photons where they have been'', but let me quote here the concluding remarks:
\begin{quote}
The photons themselves tell us where they have been. And the story they tell  is surprising. The photons do not always follow continuous trajectories. Some of them have been inside the nested interferometer (otherwise they could not have known the frequencies $f_A, f_B$), but they never entered the nested interferometer since otherwise they could not avoid the imprints of frequencies $f_E$ and $f_F$ of mirrors $E$ and $F$ leading the photons in and out of the interferometer. Only the description with both forward and backward evolving quantum states provides a simple and intuitive picture of pre- and postselected particles.
\end{quote}
The article promotes a hypothesis that trajectories of photons can be discontinuous, which is explained by the two-state vector formalism (TSVF) in which we overlap the state evolved forward from the source with a backward-evolved state reaching the detector. Being selected for a {\em Viewpoint} in {\em Physical Review Letters}, it received a lot of attention, ranging from critical comments \cite{A00,A01,A02,A03,A04, A05, A06,A07,A08,A09,A10,A11} to faithful follow-ups \cite{F01,F02,F12,F03,F04,F05,F06,F07,F08,F08,F08,F09,F10,F11} (the Reader is kindly asked to notice the comments and replies to these works).

The aim of this work is simple. Using the elementary formalism of interferometry I will confirm the basic observations of Danan, Farfunik, Bar-Ad, and Vaidman. However, later their conclusions will be confronted with somewhat deeper (though, still very basic) analysis.

One should note however, that it is not possible to exactly reproduce the interferometer from Ref. \cite{VAIDMAN}, as it lacks more detailed description of the setup. Figures 1-3 therein suggest that the mirror were pulled and pushed by a piezoelectric and hence tilted with respect to some axis. This rather causes a beam displacement, but since no dimensions are given, it is impossible to consider all possible effects. I would rather focus on systems of mirrors, which do not displace the beam, but simply apply an extra phase shift. Such a system can be easily arranged, e.g., out of a single mirror and a rooftop table. Agreeably, the correctness of a physical theory cannot depend on specifics of a mirror being used. 

Immediately, another problem arises. The amplitudes of mirror vibrations are given as about $1.5\times 10^{-7}$ rad, to be compared with the width of the beam being close to $3.7\times 10^{-4}$ rad. The idea is that the frequency marker does not disturb the the interference. Any experimentalist will confirm that settings of a perfectly tuned interferometer are a measure-zero subset of all possible settings and is unattainable in practice. In the analysis one must include the effects of marking photons on the interference pattern. Note that these small beam displacements are expected to generate prominent effects. Also, the results in Ref. \cite{VAIDMAN} (See Figure 3 therein), are presented only as small insets, using only one fifth of the axis. No effort is taken towards studying more subtle structures. As I shall show below, extremely detailed analysis of experimental data is required. The question of the magnitudes of different effects will play a crucial role below.

The amplitude of a continuous beam of light (or, equivalently, of the probability of detecting a single spontaneously emitted photon, pulsed light may cause an additional interplay between frequencies) is (proportional to) the second component, $(V(t))_2$, of vector
\begin{eqnarray}
\label{interf}
V(t)&=&\left(\begin{array}{ccc}1&0&0\\0&x&x\\0&x&-x\end{array}\right)\left(\begin{array}{ccc}1&0&0\\0&e^{i\phi_F(t)}&0\\0&0&1\end{array}\right)\nonumber\\
&\times&\left(\begin{array}{ccc}x&x&0\\x&-x&0\\0&0&1\end{array}\right)\left(\begin{array}{ccc}e^{i\phi_A(t)}&0&0\\0&e^{i\phi_B(t)}&0\\0&0&1\end{array}\right)\nonumber\\
&\times&\left(\begin{array}{ccc}x&x&0\\x&-x&0\\0&0&1\end{array}\right)\left(\begin{array}{ccc}e^{i\phi_E(t)}&0&0\\0&1&0\\0&0&e^{i\phi_C(t)}\end{array}\right)\nonumber\\
&\times&\left(\begin{array}{ccc}x&0&x\\0&1&0\\x&0&-x\end{array}\right)\left(\begin{array}{c}1\\0\\0\end{array}\right)
\end{eqnarray}
50:50 beamsplitters are used ($x=\frac{1}{\sqrt{2}}$), which is not relevant. I take $\phi_X(t)=A_0\sin 2\pi f_Xt$, with amplitude $A_0=\pi/100$, and $f_A=37,f_B=41,f_C=43,f_E=159,$ and $f_F=179$. For simplicity, these frequencies are assumed to be orders of magnitudes lower than the frequency of light. Other choices of $A_0$ change the magnitude of effects described below, but the general idea would remain the same.

 Let us read Eq. (\ref{interf}) chronologically (cf. Fig 3 in Ref. \cite{VAIDMAN}). Light is injected through a single input and immediately split into two arms. In the lower, it reaches mirror $C$ and an oscillating phase $\phi_C(t)$ is applied. In the upper arm, light acquires phase $\phi_E(t)$ and is redirected to the small loop. It is then split and can acquire either $\phi_A(t)$ or $\phi_B(t)$. The light from the small MZI is then recombined. Had it gone rightwards, it would then bounce off mirror $F$ and be recombined with the beam from the lower path. However, at $t=0$ the whole light from the small loop is directed downwards. A simple algebra leads to:
\begin{eqnarray}
\label{Ft}
(V(t))_2&=&\frac{1}{4}\left(2e^{iA_0\sin(2\pi f_Ct)}\right.\nonumber\\
&+&e^{iA_0(\sin(2\pi f_At)+\sin(2\pi f_Et)+\sin(2\pi f_Ft))}\nonumber\\
&-&\left.e^{iA_0(\sin(2\pi f_Bt)+\sin(2\pi f_Et)+\sin(2\pi f_Ft))}\right).
\end{eqnarray}
One straightforwardly recognizes terms contributed by each path in the setup. Note that due to the destructive interference in the upper arm, the two last terms, emerging from passing through the small loop, have the opposite signs, but as the small MZI is meant to be tuned, they have equal weights. Since $A_0\ll 1$, the exponents will always close to 1.

The power spectrum is obtained as
\begin{equation}
G(f)=\left|\int_0^1 e^{-2\pi i f t}(V(t))_2dt\right|^2,
\end{equation}
And since $(V(t))_2$ is a periodic function, only the integer values of $f$ are meaningful.

\begin{figure}[!h]
	\centering
		\includegraphics[width=6cm]{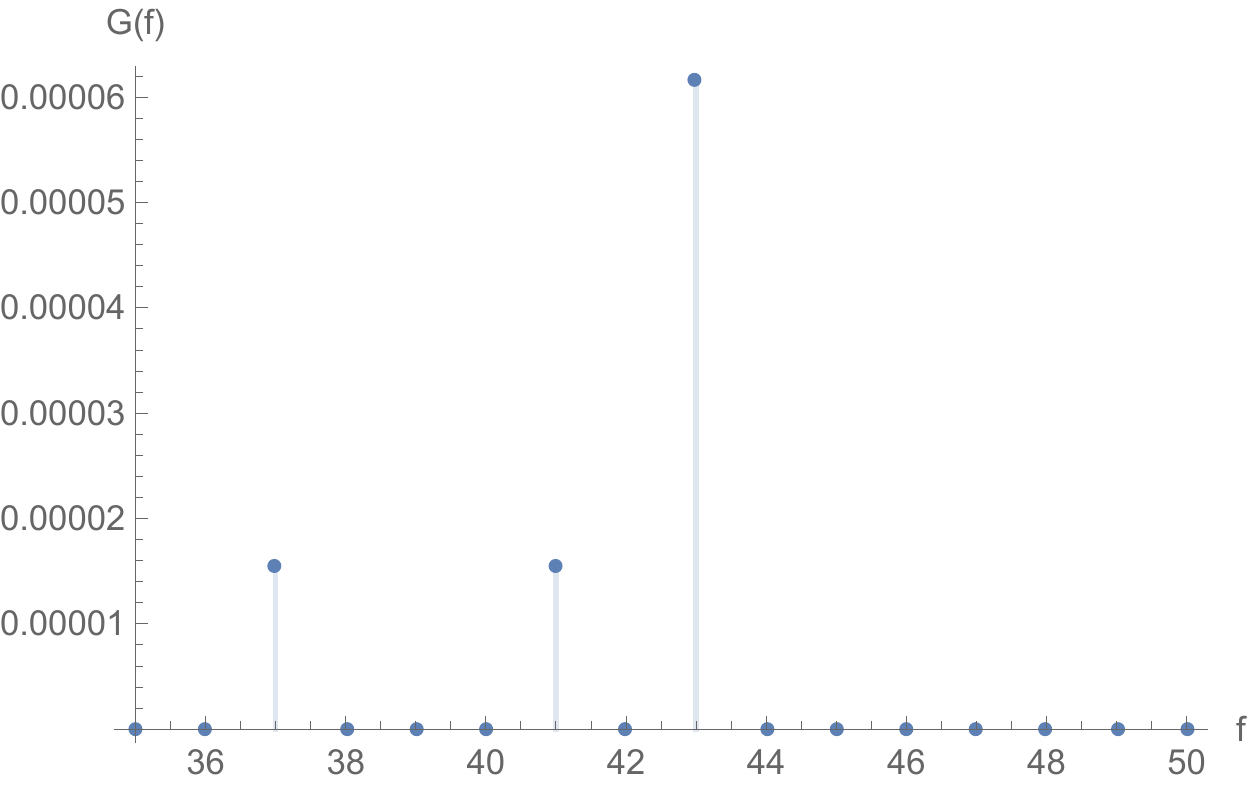}
	\caption{The power spectrum $G(f)$ in function of frequency $f$ in band $35-50$. Prominent peaks at 37 ($f_A$), 41 ($f_B$), and 43 ($f_C$) are visible.}
	\label{Fig1}
\end{figure}

\begin{figure}[!h]
	\centering
		\includegraphics[width=6cm]{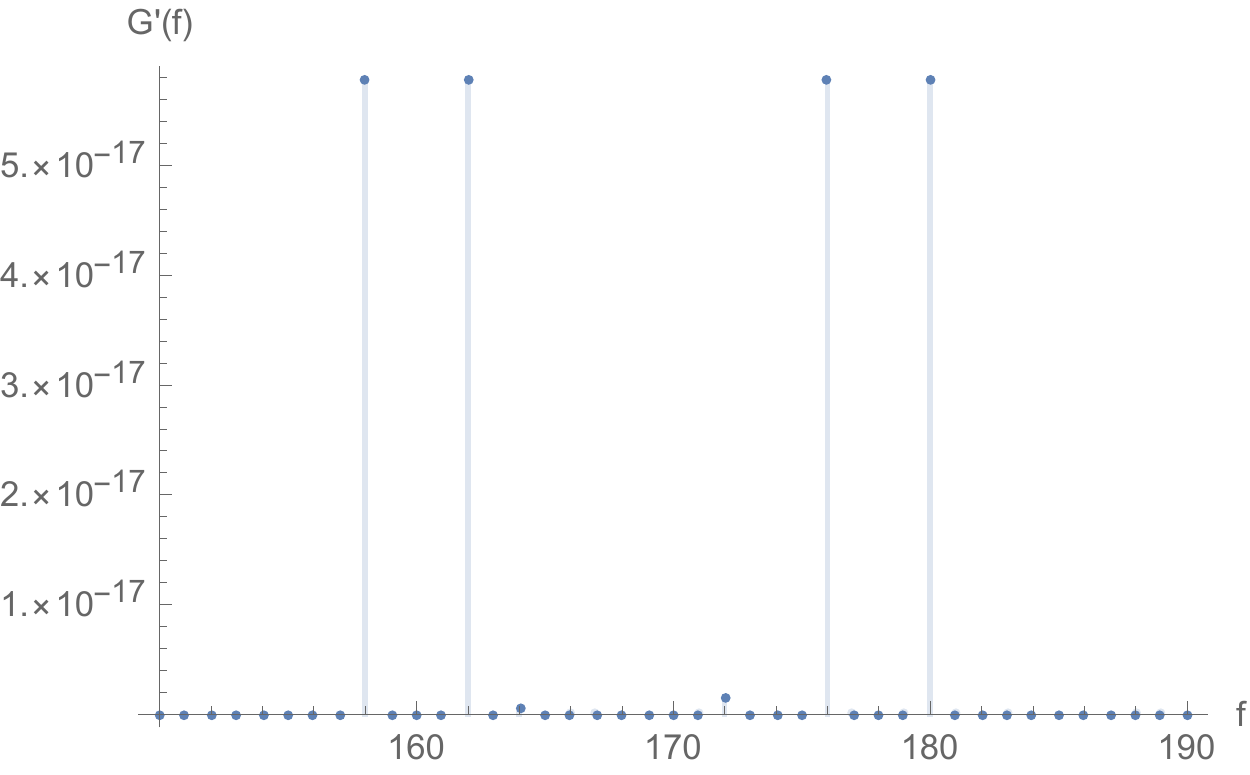}
	\caption{The power spectrum $G(f)$ in function of frequency $f$ in band $150-190$ containing $f_E$ and $f_F$. No peaks significantly above the numerical precision.}
	\label{Fig2}
\end{figure}

Figure \ref{Fig1} presents the power spectrum in the bound containing $f_A, f_B,$ and $f_C$. Indeed, we see prominent peaks at $f=37,41,43$. They are of different heights due to beam splitters being balanced. Figure \ref{Fig2} is $G(f)$ for $150\leq f\leq 190$. We see some very small peaks of an equal height at 158,162, 176 and 180, and few smaller ones elsewhere, but not at $f_E=159$ or $f_F=179$. These peaks could possibly be attributed to numerical errors, as they lie on boundary of they numerical precision of a highly oscillatory integrand in Wolfram$^{\text \textregistered}$ Mathematica$^{\text \textcopyright}$. We can proclaim the results of Danan and co-authors to be confirmed.

However, one should not draw a conclusion about radio communication in ancient Rome basing only on the lack of telegraphic poles. A necessary condition to argue for discontinuity in propagation of photons is that the power spectrum has no syndrome, which can be directly related to $f_E$ and $f_F$. Indeed, we find peeks at $f=196=f_A+f_E$, $200=f_B+f_E$, $216=f_A+f_F$, $220=f_B+f_F$. They are orders of magnitude smaller than peaks at $f_A,f_B,f_C$, but still much above the precision of calculations. Other choices of $f_E$ and $f_F$ reveal analogous effects. Let me point out that peaks at combinations of frequencies were already mentioned in Ref. \cite{A04}.

\begin{figure}[!h]
	\centering
		\includegraphics[width=6cm]{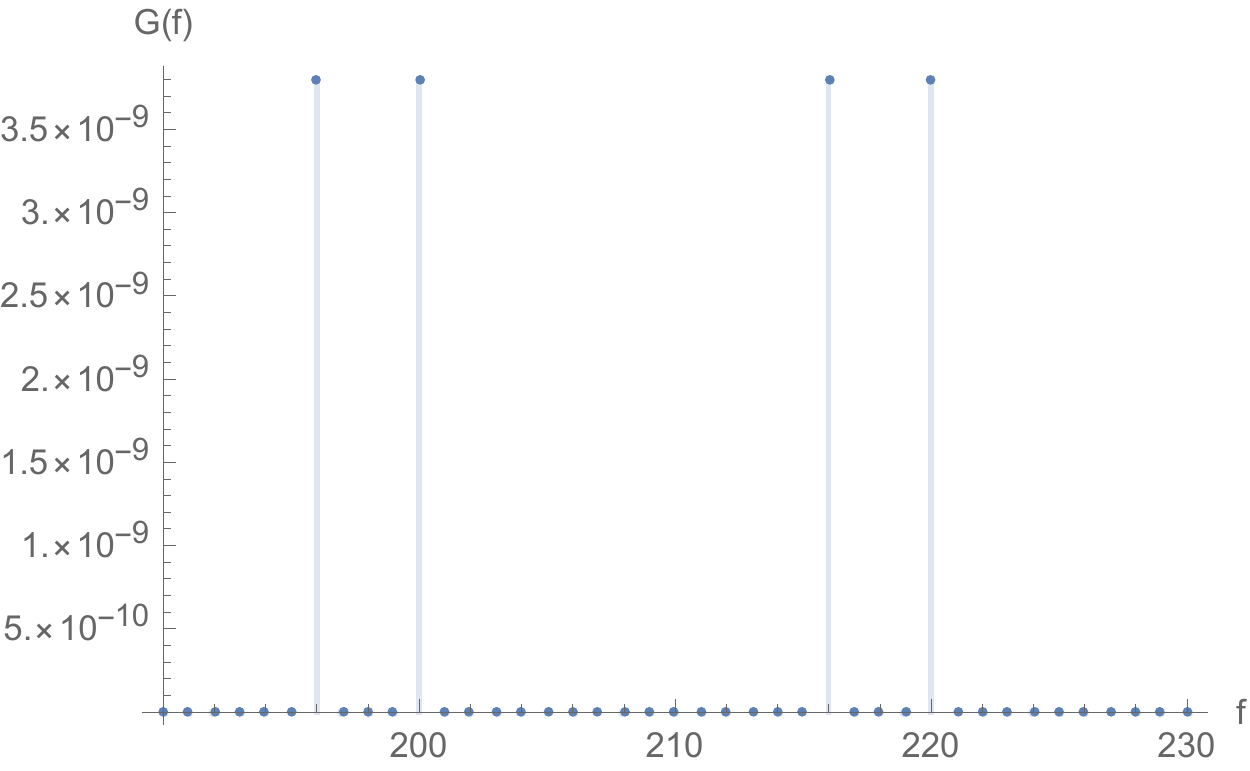}
	\caption{The power spectrum $G(f)$ in function of frequency $f$ in band $190-230$ with peaks at 196, 200, 216 and 220.}
	\label{Fig3}
\end{figure}

Let me now demonstrate analytically the existence of the peaks. To do this, one simply uses the Taylor expansion of the exponent, $e^x=\sum_{n=0}^\infty\frac{x^n}{n!}$, and subsequently expands the powers of its arguments, $(a+b)^n=\sum_{j=0}^n\left(\begin{array}{c}n\\j\end{array}\right)a^jb^{n-j}$. Because the two last terms of $(V(t))_2$ have the opposite signs, all terms, which do not depend on $f_A, f_B$ or $f_C$ (except for the zeroth term of $e^{2\pi i f_A t}$), vanish, which explains the lack of peaks at $f_E$ and $f_F$. We then use identity $\sin a\sin b=\frac{1}{2}(\cos(a-b)-\cos(a+b))$ to see the peaks at $f_{A/B}\pm f_{E/F}$ with magnitudes of the order of $A_0^2$ Likewise, we find peaks at $n_1 f_B+n_2 f_E+n_3 f_F, n_1 f_B+n_2 f_E+n_3 f_F$ and $n_4 f_C$ (including one of those mentioned in Ref. \cite{A03}) with magnitudes of orders $A_0^{n_1+n_2+n_3}$ and $A_0^{n_4}$, respectively, where $n_1\neq 0, n_2, n_3,n_4$ are integers.

\begin{figure}[!h]
	\centering
		\includegraphics[width=6cm]{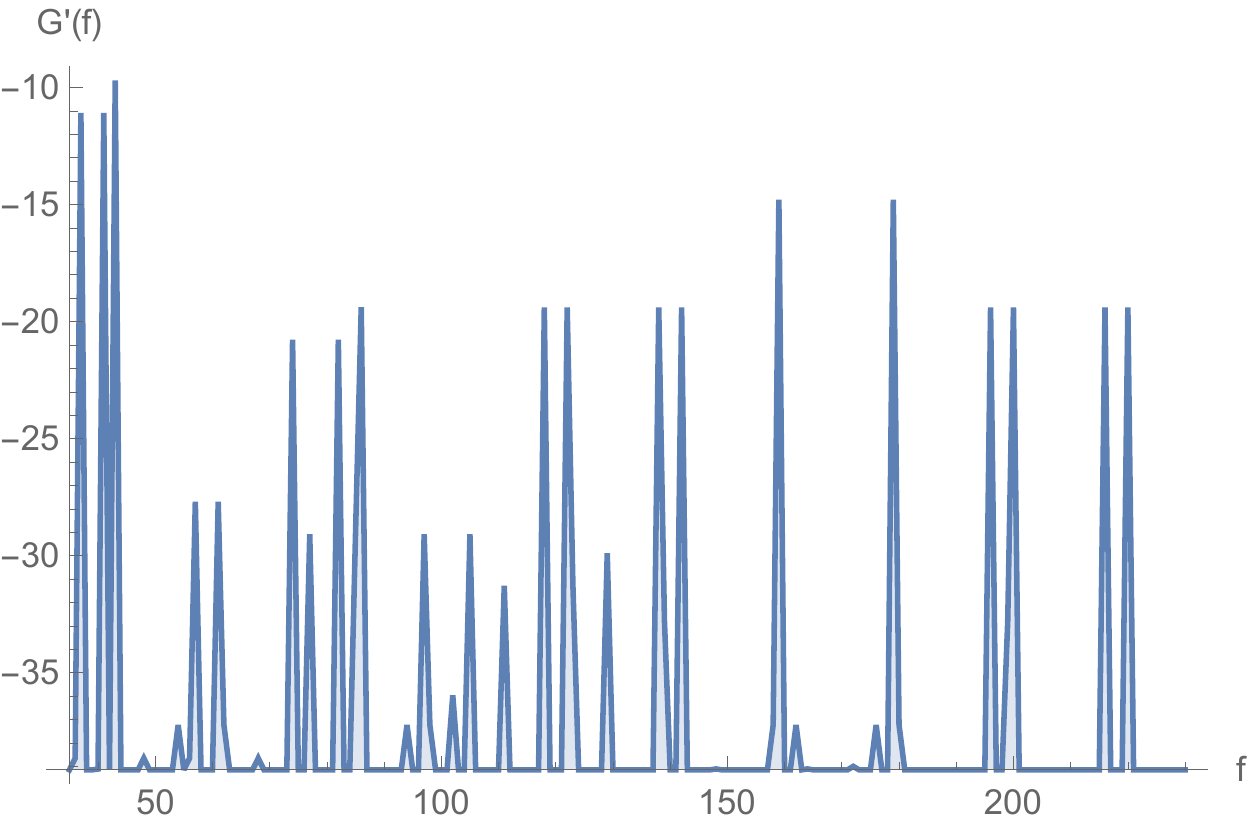}
	\caption{The logarithm of the power spectrum $\ln(G'(f)+10^{-17})$ for the detuned interferometer in function of frequency $f$ in band $35-230$ with peaks at 159 and 179.}
	\label{Fig4}
\end{figure}

If this is not a convincing argument, consider some instability of the small loop. Imagine that behind mirror B, but in front of the beam splitter recombining the paths of the small loops, there is an extra delay applying a phase factor. This makes two last terms in $(V(t))_2$  significantly different, and the light can ``officially'' reach the detector by partially constructive interference:
\begin{eqnarray}
\label{Fpt}
(V'(t))_2&=&\frac{1}{4}\left(2e^{iA_0\sin(2\pi f_Ct)}\right.\nonumber\\
&+&e^{iA_0(\sin(2\pi f_At)+\sin(2\pi f_Et)+\sin(2\pi f_Ft))}\nonumber\\
&-&\left.e^{iA_0(\sin(2\pi f_Bt)+\sin(2\pi f_Et)+\sin(2\pi f_Ft))+i\pi/20}\right),\nonumber\\
G'(f)&=&\left|\int_0^1 e^{-2\pi i f t}(V'(t))_2dt\right|^2.
\end{eqnarray}
 In such a situation, we directly see peaks corresponding to $f_E$ and $f_F$, see Figure \ref{Fig4}. This extra phase shift can be, however, introduced after a photon would enter the small MZI, as in Wheeler's delayed choice experiment \cite{WHEELER}. How can then light decide whether to manifest the effects of encountering mirror $E$? The hypothesis of discontinuous trajectories of a photon lacks the logic of ``choosing'' locations where it ``materializes''.

What about the triumph of TSVF? I have shown that it was brought up to explain an effect that was never there. Indeed, the two-state-vector formalism, as presented in \cite{VAIDMAN} will not reconstruct all the features of the discussed experiment. It can hence, at most, play a role of a toy model, similar to the spinless Bohr's model of an atom. It allows to explain the most basic features (quantum numbers), but ignores others (Zeeman effect), and may introduce false notions (localized electron orbits). However, TSVF introduces an extra entity, namely the back-evolved state, and is eventually more demanding computationally. Other possible weaknesses of TSVF are mentioned in Ref. \cite{A04}.

In conclusion, the most basic interferometric calculations fully explain the observations presented in Ref. \cite{VAIDMAN} without a need for TSVF, backward-evolving states, ``secondary presence'' \cite{F12}, or weak measurements/values/traces. They are simply effects of the destructive interferences. The experiment certainly cannot be explained by the means of a classical particle (not being in a spatial superposition), but it confirms the interaction with all optical elements in the setup. However, insufficient presentation and analysis of results has led the Authors of Ref. \cite{VAIDMAN} to premature and false conclusions. 

The most surprising thing about Ref. \cite{VAIDMAN} is its reception in the community. A number of us \cite{F01,F02,F12,F03,F04,F05,F06,F07,F08,F09,F10,F11} have questioned a well confirmed, century-old theory for a special case of an insufficiently analyzed experiment. Meanwhile, the effects discussed in Ref. \cite{VAIDMAN} are precisely predicted by standard quantum mechanics, together with other, more subtle, consequences that need to be included.

The key point in the reasoning of Danan, Farfurnik, Bar-Ad, and Vaidman was to convince the Reader that bouncing from a vibrating mirror imprints this frequency in the spectrum of light. This was easily obtained with Fig. 1 in the Reference and the text thereabout. However, It was not mentioned that such imprints can accumulate resulting in peaks at combinations of the used frequencies, or that the spectrum power function can be highly involved due to interference.

This work was supported by the Polish National Center for Science (NCN) Grant No. UMO-2015/19/B/ST2/01999 (task 1). 


\begin{thebibliography}{99}
\bibitem{YOUNG}  T. Young, {\em A Course of Lectures on Natural Philosophy and the Mechanical Arts} {\bf 1}. {\em Lecture} {\bf 39}, pp. 463–464 (1807).
\bibitem{MZI} L.Zehnder, {\em Ludwig Zeit. für Instr.} {\bf 11}, pp. 275–285 (181); L. Mach, {\em Zeitschrift für Instrumentenkunde} {\bf 12}, pp. 89–93 (1892).
\bibitem{WHEELER} J. A. Wheeler, in {\em Quantum Theory and Measurement} pp. 182-213 (Princeton University Press, 1984).
\bibitem{ASPECT} V. Jaques, {\em et al.}, {\em Science} {\bf 315}, 966-968 (2007). 
\bibitem{VAID1} L. Vaidman, {\em Phys. Rev. A} {\bf 87}, 052104 (2013).
\bibitem{VAIDMAN} A. Danan, D. Farfurnik, S. Bar-Ad, and L. Vaidman, {\em Phys. Rev. Lett.} {\bf 111}, 240402 (2013).
\bibitem{A00}  H. Salih, {\em Front. Phys.} {\bf 3}, 47 (2015).
\bibitem{A01} P. Saldanha, {\em Phys. Rev. A} {\bf 89} 033825 (2014).
\bibitem{A02} B. Svensson, e-print arXiv:1402.4315 [quant-ph] @ arxiv.org (2014).
\bibitem{A03} J.-H. Huang {\em et al.}, e-print arXiv:1402.4581 [quant-ph] @ arxiv.org (2014).
\bibitem{A04} M. Wie\'sniak, e-print arXiv:1407.1739 [quant-ph] @ arxiv.org (2014).
\bibitem{A05} D. A. Slavov, {\em  Phys. Part. Nucl.} {\bf 46}, 665 (2015).
\bibitem{A06} K. Bartkiewicz {\em et al.}, {\em Phys. Rev. A} {\bf 91}, 012103 (2015).
\bibitem{A07} R. B. Griffith, {\em Phys. Rev. A.} {\bf 94}, 032115 (2016).
\bibitem{A08} M. Bula, {\em et al.}, {\em Phys. Rev. A} {\bf 94}, 052106 (2017).
\bibitem{A09} G. N. Nikolaev, {\em JEPT Lett.} {\bf 105}, 152 (2017).
\bibitem{A10} B.-G. Englert {\em et al.}, e-print arXiv:1704.03722 [quant-ph] @ arxiv.org (2017).
\bibitem{A11} D. Sokolovski, {\em Phys. Lett. A} {\bf 381}, 227 (2017).
\bibitem{F01} J. Lundeen, {\em Physics} {\bf 6}, 133 (2013).
\bibitem{F02} L. Vaidman, {\em Quantum Stud.: Math. Found.} {\bf 1}, 5 (2014).
\bibitem{F12} L. Vaidman, {\em Phys. Rev. A} {\bf 89}, 024102 (20174).
\bibitem{F03} L. Fu, F. A. Hashimi, Zh. Jun-Xiang, and Zh. Shi-Yao, {\em Chin. Phys. Lett.} {\bf 32}, 050303 (2015).
\bibitem{F04} E. Cohen and C. Elitzur, {Jour. Phys.: Conf. Series} {\bf 626}, 012013 (2015).
\bibitem{F05} D. Puhlmann {\em et al.}, {\em Phys. Scr.} {\bf 91}, 023006 (2016).
\bibitem{F06} A. Kellerer, S. Wright, and S. Lacour, {\em Am. Jour. Phys.} {\bf 85}, 6 (2017).
\bibitem{F07} A. Ben-Israel {\em et al.}, {\em Chin. Phys. Lett.} {\bf 34}, 020301 (2017).
\bibitem{F08} Z.-Q. Zhou {\em et al.}, {\em Phys. Rev. A} {\bf 95}, 042121 (2017).
\bibitem{F09} Q. Duprey and A. Matzkin {\em Phys. Rev. A} {\bf 95}, 032110 (2017).
\bibitem{F10} Y. Aharonov, E. Cohen, A. Landau, A. C. Elitzur {\em Sci. Rep.} {\bf 7}, 351 (2017).
\bibitem{F11} B. de Lima Bernardo, A. Canabarro, and S. Azevedo {\em Sci. Rep.} {\bf 7}, 39757 (2017).
\end{thebibliography}
\end{document}